\documentclass[aps,pra,twocolumn,showpacs,amsmath,amssymb,superscriptaddress]{revtex4-2}

\usepackage{graphicx}
\usepackage{color}
\usepackage{epstopdf}
\usepackage{listings}
\usepackage{braket}
\usepackage{float}
\usepackage{hyperref}

\begin{document}
\title{Formation and stability of vortex  solitons in nematic liquid crystals}

\author{Pawel S. Jung}
\thanks{Corresponding author: pawel.jung@ucf.edu}
\affiliation{CREOL, The College of Optics and Photonics, University of Central Florida, Orlando, Florida 32816, USA}
\affiliation{Faculty of Physics, Warsaw University of Technology, Warsaw, Poland}
\author{Yana V. Izdebskaya}
\affiliation{Laser Physics Centre, Research School of Physics,  The Australian National University, Canberra ACT 0200, Australia}
\affiliation{Nonlinear Physics Centre, Research School of Physics,  The Australian National University, Canberra ACT 0200, Australia}
\author{Vladlen G. Shvedov}
\affiliation{Laser Physics Centre, Research School of Physics,  The Australian National University, Canberra ACT 0200, Australia}
\author{Demetrios N. Christodoulides}
\affiliation{CREOL, The College of Optics and Photonics, University of Central Florida, Orlando, Florida 32816, USA}
\author{Wieslaw Krolikowski}
\affiliation{Laser Physics Centre, Research School of Physics,  The Australian National University, Canberra ACT 0200, Australia}
\affiliation{Science Program, Texas A \& M University at Qatar,  Doha, Qatar}


\begin{abstract}
We study the propagation dynamics of bright optical vortex solitons in nematic liquid crystals with a nonlocal reorientational nonlinear response. We investigate the role of optical birefringence on the stability of these solitons. In agreement with recent experimental observations, we show that the birefringence-induced astigmatism can eventually destabilize these vortex solitons. However, for low and moderate birefringence, vortex solitons can propagate stably over experimentally relevant distances.
\end{abstract}

\maketitle


Vortex solitons represent non-diffracting singular bright rings propagating in nonlinear media by properly balancing diffraction and self-focusing nonlinear effects \cite{Kruglov:pla:85,Denschlag:science:00,Desyatnikov:progopt:05}. 
They are rather elusive objects as they tend to break-up during evolution. In typical homogenous local nonlinear media such as those exhibiting Kerr-like and saturable nonlinearities, vortex solitons are subject to azimuthal instabilities that can break them into individual bright solitons~\cite{Soto:pra:91,Malomed:joptb:05}. A few approaches have been proposed to arrest this instability. They involve, for instance, use of vector solitons formed by the incoherent superposition of a vortex component and a fundamental beam ~\cite{Yang:pre:03,Izdebskaya:ol:15} or two vortices with opposite charges ~\cite{Izdebskaya:ol:12}. In addition, systems with nonlinear gain and loss may support stable vortex solitons~\cite{Porras:pra:16}.

As far as single-component vortex solitons in passive nonlinear media are concerned, their azimuthal instability can be prevented by the simultaneous action of competing nonlinearities. In this nonlinear model, a substantial defocusing contribution to the otherwise focusing nonlinearity may stabilize the vortex soliton as has been predicted for the so-called cubic-quintic model ~\cite{Quiroga:josab;97}. This model has been later employed to theoretically demonstrate stabilization of even higher dimensional vortices~\cite{Mihalache:prl:02}.  
 
  \begin{figure}[h]
\centerline{\includegraphics[width=1.0\columnwidth]{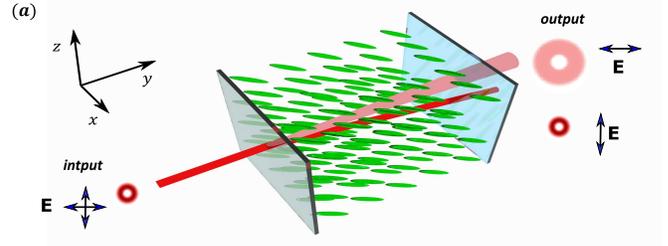}}%
\caption{\label{fig_1}  Schematic representation of beam propagation in a nematic liquid crystal cell. The uniform molecular orientation defines the ordinary (x-y) and extraordinary planes (y-z). An arbitrarily polarized input beam splits into an ordinary (freely diffracting) and an extraordinary (self-focusing) component.}
\end{figure} 
 
 On the other hand, it appears that the spatially nonlocal nonlinearity  may stabilize  vortex solitons. Nonlocality creates a broad refractive index structure that serves as a waveguide  (or trapping potential) and may prevent a soliton from breaking up~\cite{Briedis:opex:05,Ramaniuk:oc:19}. 
The spatial nonlocality of nonlinearity is commonly associated with certain transport processes such as heat transfer in thermal media ~\cite{Dabby:apl:68,Kwasny:ol:20}, diffusion of atoms or molecules ~\cite{Suter:pra:93}. 
It is also natural for media with long-range inter-molecular or atomic interactions, such as in cold atomic gases ~\cite{Dalfovo:rmp:99}  or liquid crystals (LC) ~\cite{Peccianti:pr:12}.  The latter is particularly important  because their large noninstantaneous  orientational  and spatially nonlocal nonlinearity and strong electro-optical response  allows one to realize  various types of  solitons, the so-called nematicons.

The nonlinear response of LC is caused by the light-induced reorientation of their constituent molecules and subsequent increase of the extraordinary refractive index. To this end, the molecules of liquid crystals must be externally pre-oriented. It is commonly done by placing the liquid crystal between two planar boundaries, which enforce initial molecular orientation by either applying an external AC voltage ~\cite{Peccianti:apl:00} or rubbing the surfaces, which pre-orients the molecules.   Alternatively, one may use an external magnetic field, which also orients molecules as they exhibit strong diamagnetism~\cite{Yana:ol:18}. Until recently,  the vortex-like solitons, which were observed in liquid crystal cells, were vector vortex solitons, i.e., structures formed by a co-propagating vortex beam with another bell-like beam which served as a guide for the vortex component ~\cite{Izdebskaya:ol:15,Izdebskaya:jo:16}.

However, formation and stable propagation of real (single beam, or scalar) vortex solitons have been reported very recently in two experimental studies. Izdebskaya {\em et al.} ~\cite{Yana:ol:18} employed a magnetic field for controlling the pre-orientation and the degree of nonlocality~\cite{Izdebskaya:nc:17}, while Laudyn {\em et al.} ~\cite{Laudyn:oe:20} used the mechanically conditioned liquid crystal cell's surfaces to enforce the proper initial molecular orientation.
The latter work demonstrated that the vortices' stability critically depends on the optical anisotropy of the liquid crystal. It was shown that liquid crystals with low birefringence do allow the formation and propagation of vortex solitons. High birefringence introduces strong astigmatism, which, subsequently, destabilizes the vortex soliton, which quickly disintegrates during propagation.

In this work, we address the stability of vortex solitons in nematic liquid crystals with a molecular orientation-based nonlinearity.
In particular, we show how this stability depends on the birefringence induced astigmatism. Our results provide a detailed explanation of vortex instability in nematic liquid crystals and hence complement recent experimental and earlier theoretical works on this subject.

We consider the propagation of optical beams in the bulk of a nematic liquid crystal having a uniform molecular initial orientation, as depicted in Fig.1. The initial orientation can be achieved using an external low-frequency AC electric or DC magnetic field.  In general, the LC represents a uniaxial optical medium. Therefore an arbitrary polarized optical beam that is incident onto the LC cell will split into an ordinary and an extraordinary component. The former propagates linearly, experiencing diffraction. The latter may induce reorientation of molecules, thus increasing the effective refractive index if its intensity is high, and, consequently, could experience self-focusing in which case it will propagate as a spatial optical soliton.
 For simplicity and without any loss of generality, in what follows,  we will use a simplified description of beam evolution in a LC, where we treat the medium as a uniaxial optical crystal. Assuming that the input beam is extraordinarily polarized, the spatial evolution of the electric field's amplitude $E(x,z)$, propagating predominantly along the y-axis (neglecting absorption), is described by the following wave equation \cite{Fleck:83,alberucci2010propagation}:

\begin{eqnarray}\label{nls}
 && 2ik_{o}n(\theta_{o})\left( \frac{\partial E}{\partial y} + \tan\delta(\theta)\frac{\partial E}{\partial z} \right)+\frac{\partial^{2} E}{\partial x^{2}}
+ D_{z}\left(\theta\right)\frac{\partial^{2} E}{\partial z^{2}} \nonumber \\& &+k_{o}^{2}\left( n^{2}(\theta)-n^{2}(\theta_{o}) \right)E=0,
\end{eqnarray}
\noindent where $ k_o$=2$\pi/\lambda_o$, $\delta\left(\theta\right)$ is the walk-off angle  calculated along the beam axis, $\theta_o$ is the initial  molecular orientation (in the absence of light), $D_z=\cos^{2}\theta+\gamma^2\sin^{2}\theta$ is the diffraction coefficient across $z$  and 
$n\left(\theta\right) = \left(\cos^{2}\theta/n_{o}^{2}+\sin^{2}\theta/n^{2}_{e}\right)^{-1/2} $
is an effective index of refraction for the y-polarized (i.e., extraordinary) light. Here $n_o$ and $n_e$ represents ordinary and extraordinary refractive indices, respectively and 
$\gamma^2=n_e^2/n_o^2$.
The index $n(\theta)$ and  walk-off angle $\delta$~\cite{Piccardi:apl:10} depend on the local orientation of molecules $\theta$ which is governed by the following relation~\cite{Assanto:jqe:03}:
\begin{equation}\label{or}
 \frac{\partial^{2} \theta }{\partial z^{2}} +\frac{\partial^{2} \theta }{\partial x^{2}} 
+ \frac{\Delta\varepsilon \varepsilon_{o}}{2K}|{E}|^{2} \sin 2\theta
 =0. 
\end{equation}
In Eq.(2) $K$ is an effective elastic constant  and $\Delta\varepsilon= n_e^2-n_o^2$ is  the dielectric ~\cite{Khoo:jqe:93} anisotropy of the liquid crystal.  

In what follows, we will neglect the walk-off effect (as defined by $\delta$) as it does not play a role in the vortex soliton instability. A crucial aspect associated with Eq.(1) is the presence of anisotropy in the Laplacian, reflected by an effective an diffraction coefficient $D_z(\theta)$. This term reflects an anisotropy in the diffraction along the extraordinary and ordinary axes. In effect, it is a source of astigmatism which affects beam propagation, and as we will show later, it is directly responsible for the instability of the vortex soliton. The diffraction coefficient is a function of the LC's optical anisotropy (if $D_z=1$ the medium is isotropic).  
 Fig.2(a) depicts the diffraction coefficient as a function of the initial molecular orientation ($\theta_o$) and an optical  anisotropy ($\Delta n=n_e-n_o$). We used values ranging from $\Delta n=0.1$ to $\Delta n=0.4$ , which correspond to widely available types of nematic liquid crystals. The lack of cylindrical symmetry in Eq.(1) directly affects the optical beam's propagation even in the linear regime. 
 
\begin{figure}[h]
\centerline{\includegraphics[width=1.0\columnwidth]{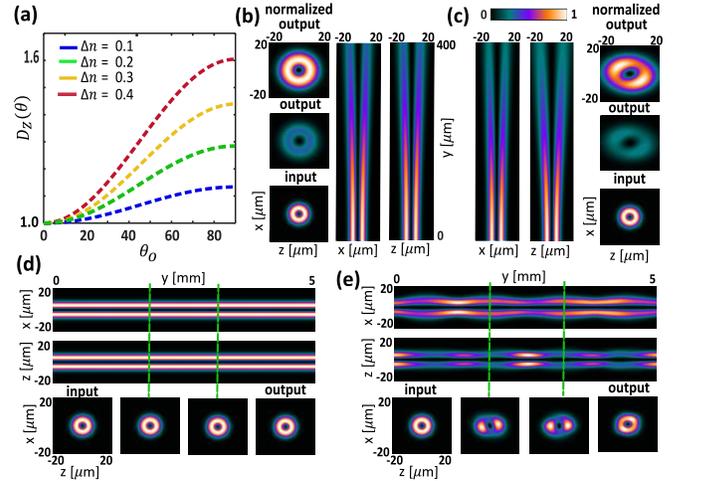}}
\caption{\label{fig_1} (a) The effective diffraction coefficient along $z'$-direction for few values of optical anisotropy. (b-c) 
Illustrating the effect of optical anisotropy on propagation of vortex beams with charge $m=1$ and waist $w_o=6 {\mu}$m  in liquid crystal ($\lambda=0.8$ $\mu m$) (b,c) linear regime (low intensity), (d,e) nonlinear regime (high intensity). $D_{z}=1$ in (b,d) and $D_{z}=1.6$ in (c,e).}
\end{figure}

This effect is illustrated in Figs.2(b-c), which depict the propagation of a single charge optical vortex beam with (b) and without (c) anisotropy.  In the latter case, the vortex retains its the structure while gradually diffracting. However, in the anisotropic medium, ensued astigmatism completely breaks down the vortex beam. To better understand the effect of optical anisotropy and initial molecular orientation of LC on the stability of vortex solitons, we conducted a series of numerical simulations by varying the strength of anisotropy ($\Delta n$) and initial molecular orientation ($\theta_o$). 
In all subsequent numerical simulations, we use a Laguerre-Gaussian-like beam as an initial condition.  The beam size was kept at  6$\mu$m. The input power was chosen such that the beam would constitute an  exact vortex soliton solution in the isotropic case ($D_z=1$). Consequently,  the input power was adjusted to account for the reorientation nonlinearity dependence on   $\theta_0$,  reaching its maximum near $\theta_0 \approx 45^{\circ}$.
In Fig.2 (e), we show the nonlinear propagation of a single charge vortex soliton in the presence of strong anisotropy. Clearly, the soliton becomes quickly unstable. For comparison, in Fig.2(d), we show the same soliton dynamics without an anisotropy, which as expected is stable. 

\begin{figure}[h]
\centerline{\includegraphics[width=1.0\columnwidth]{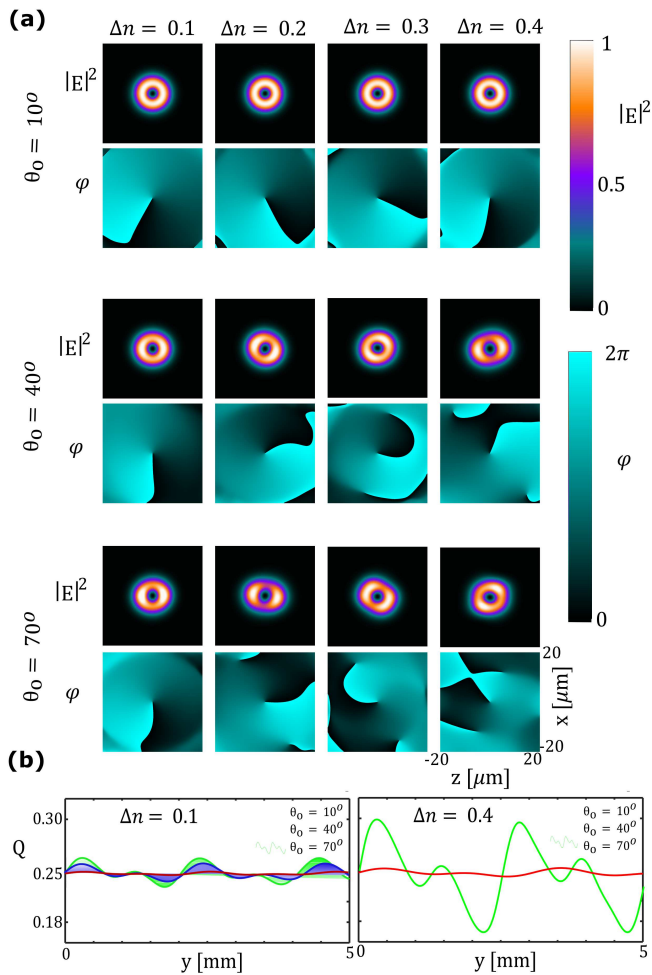}}%
\caption{\label{fig_1} 
Illustrating the effect of optical anisotropy on the vortex solitons structural instability having charge 1 in nematic liquid crystals with various degrees of anisotropicity. (a) Beam intensity ($|E|^2$) and phase ($\varphi$)  distribution at the output z=0.5cm.  (b) The asymmetry coefficient {\em{Q}} as a function of distance for two values of anisotropy. }
\end{figure}

In Fig.3, we summarize the results of numerical simulations of a single charge vortex soliton propagating in the presence of an optical anisotropy. These plots depict the output intensity distribution of an initially exact vortex soliton, after 0.5 cm propagation distance in a LC (Fig.3(b)). Each row corresponds to a fixed initial molecular orientation while each column to a specific value of optical anisotropy, ranging from low to high. The vortex soliton instability is quantified in Fig.3(b) which  depicts the asymmetry coefficient $Q=\left(\iint_{z,x=0}^{z,x=+\infty}|E|^2dxdz\right)/\iint_{z,x=-\infty}^{z,x=+\infty}|E|^2dxdz$,  for two values of anisotropy.
The plots confirm that the larger the initial molecular orientation angle is, the faster the astigmatism-induced  vortex soliton instability develops, even for a small linear anisotropy.  On the other hand, one can achieve a relatively stable vortex soliton even in highly anisotropic crystals, provided the initial molecular orientation angle is sufficiently small.  For small orientation angles the nonlinearity is and hance a higher input power is required to maintain the soliton. However, as the plot shows, even for $\theta_0=40^{\circ}$, the charge one vortex soliton may, in principle, propagate over  5$mm$ or more. 

\begin{figure}[h]
\centerline{\includegraphics[width=1.0\columnwidth]{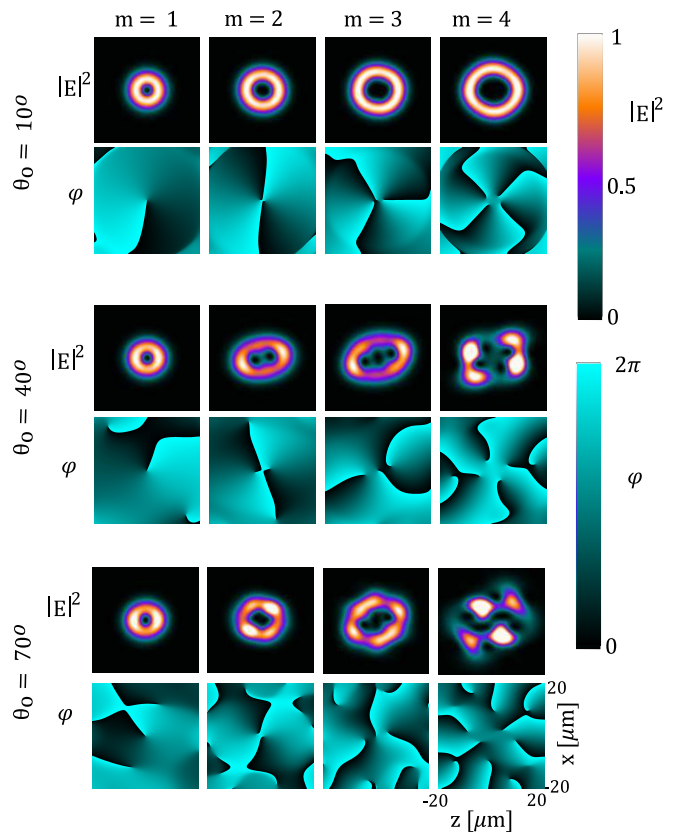}}%
\caption{\label{fig_1} 
Astigmatism-induced structural instability of a high charge  vortex solitons in  a nematic liquid crystal with a relatively large anisotropy ($\Delta n=0.4$) . In each case, graphs depict the vortex solitons intensity ($|E|^2$) and phase ($\varphi$) profiles at the same output distance $z=0.5~$cm.}
\end{figure}
The physics behind the instability of these vortex solitons can be explained by using a waveguide approach. For the vortex soliton to be stable, it must be a mode of a self-induced (via nonlinearity) waveguide. In the isotropic case, the waveguide has a cylindrical symmetry, and the vortex beam can be represented as in the quadrature superposition of two first excited (dipole-like)  modes. Degeneracy ensures that both modes propagate with the same phase velocities, thus preserving the stationary singular phase  structure. In our situation, the mode degeneracy is lifted because of the difference in the diffraction coefficients along the principal x and z' directions. Consequently, the constituent in-quadrature dipole modes travel with different phase velocities, thus destroying the initial vortex phase and intensity structures.  The larger the $D_z(\theta$), the quicker the vortex destabilizes and ultimately breaks up.    
On the other hand, it is well known that vortices do not have to exhibit circular symmetry. The so-called non-canonical vortices can display an elliptical intensity distribution and an angularly nonuniform phase gradient ~\cite{noncanonical}.  Since the LC anisotropy breaks the circular symmetry, one might expect that it could, perhaps,  support elliptical vortex solitons.  Such elliptical vortex solitons were shown to exist in Bose-Einstein condensates, trapped in anisotropic harmonic potentials \cite{Mihalache:josa:10}. In a liquid crystal, the nonlocal self-induced refractive index change plays the role of this  "external'' potential. However, because of the radial symmetry of the nonlocal response in LC, which tends to weaken the anisotropic nonlinear index change,  the elliptical vortex soliton cannot maintain its structure and breaks down, as we confirmed in additional simulations. The nonexistence of elliptic vortex solitons has been also confirmed in earlier studies of nonlocal media with anisotropic linear  or nonlinear properties~\cite{MingShen:col:14,gao2020elliptic}.
However, a co-propagating vortex beam with a fundamental bell-like beam should stabilize the former even for a strong anisotropy. In such a case, the fundamental beam plays the role of an external anisotropic potential.  Together with the nonlinear contribution from the vortex itself, it may restore  the degeneracy of the two principal orthogonal dipole-like modes in the self-induced waveguide, that supporting the vortex's stable soliton propagation.

For completeness, we also studied the behavior of high charge vortex solitons. A summary of numerical simulations for vortices with charges 1-4 is depicted in Fig.4. It is clear that these solitons quickly lose stability and breakup, first into their constituent fundamental components followed by a complete disintegration. This happens even for small initial molecular orientation angles. The breakup mechanism is similar to that seen for charge 1 vortices - lack of degeneracy between constituent modes forming high-charge vortices. In addition, there is a natural tendency for high charge vortices to split into fundamental vortices under small perturbations, even in the linear regime.   
Notice that even though the vortex soliton ultimately breaks down during propagation, its remnants stay close to the propagation axis. In this case the strong spatial LC nonlocality induces an overall waveguide-like structure that traps the original soliton and its remnants during evolution.

In conclusion, we numerically studied the structural stability of vortex solitons in nematic liquid crystals with a nonlocal reorientation nonlinearity response.  We showed that the anisotropy-induced astigmatism can destabilize the solitons and can lead to a breakup. However, the process can be slow for low anisotropies and small initial orientation angles, thus allowing almost stable propagation of red charge  vortex  solitons over an experimentally relevant distance. On the other hand, vortex solitons with higher charges quickly lose their stability and disintegrate during propagation even for moderate degrees of astigmatism. Our results explain recent experimental works on vortex solitons in low and high birefringent nematic liquid crystals. \\

Founding: The  Qatar National Research Fund (grant NPRP12S-0205-190047), the Polish National Science Center (Contract No. UMO-2013/08/A/ST3/00708) and the Polish Ministry of Science and Higher Education (1654/MOB/V/2017/0),the ONR MURI (N00014-20-1-2789),the MPS Simons collaboration (Simons grant 733682).

%


\end{document}